\documentclass[12pt,epsfig]{iopart}
\usepackage{epsfig}
\begin{document}

\newcommand{\alps}{\alpha_s}
\newcommand{\lam}{\Lambda_{\mathrm{QCD}}}
\newcommand{\deta}{\Delta \eta}
\newcommand{\pom}{{I\!\!P}}
\def\lapproxeq{\lower .7ex\hbox{$\;\stackrel{\textstyle<}{\sim}\;$}}
\def\gapproxeq{\lower .7ex\hbox{$\;\stackrel{\textstyle>}{\sim}\;$}}
\def\alpsb{\bar{\alpha}_s}
\def\half{\frac{1}{2}}

  \begin{flushright}
    CERN--TH/99--394\\
    MAN/HEP/99/6 \\
    MC-TH-99/18 \\
    December 1999
  \end{flushright}

\title[High t photons]{Diffractive production of high-$p_t$ photons at HERA}

\author{B E Cox\dag\ and J R Forshaw\ddag\footnote[3]{On leave of absence
from \dag.}}

\address{\dag\ Department of Physics and Astronomy, University of Manchester,
\\ Manchester M13 9PL, UK}

\address{\ddag\ Theory Division, CERN, 1211 Geneva 23, Switzerland}

\begin{abstract}
We study the diffractive production of high $p_t$ photons at HERA. We have
implemented the process as a new hard sub-process in the HERWIG event
generator in order to prepare the ground for a future measurement. 
\end{abstract}




\section{Introduction}
One of the cleanest of all diffractive processes is that of diffractive photon
production in the process $\gamma p \to \gamma Y$ where the photon carries a
large $p_t$ and is well separated in rapidity from the hadronic system $Y$.
The process can be measured at HERA \cite{Ivanov1,Ivanov2,Evanson}. 
The largeness of the transferred momentum 
$-t \approx p_t^2 \gg \Lambda_{{\rm QCD}}^2$ ensures the applicability of
perturbative QCD. Unlike diffractive meson production this process has the
advantage that the hard subprocess is completely calculable in perturbation
theory. The only non-perturbative component resides in the parton density
functions of the proton that factorize in the usual manner.

Theoretical interest in this process dates back to the work of 
\cite{Ginzburg} where calculations were performed in fixed order
perturbation theory and to lowest order in $\alpha_s$. Recent work has
extended this calculation to sum all leading logarithms in energy, for real
incoming photons \cite{Ivanov2} and for real and virtual incoming photons
\cite{Evanson}. The cross-section for $\gamma q \to \gamma q$ can be written
\begin{equation}
\frac{d \sigma_{\gamma q}}{dp_t^2} 
\approx \frac{1}{16 \pi \hat{s}^2} \left| A_{++} \right|^2
\end{equation}
and we have ignored a small contribution that flips the helicity of the
incoming photon. The photon-quark CM energy is given by $\hat{s}$. To leading
logarithmic accuracy \cite{Ivanov2,Evanson}
\begin{equation}
A_{++} = i \alpha \alpha_s^2 \sum_q e_q^2 \frac{\pi}{6} \frac{\hat{s}}{p_t^2}
\int_{-\infty}^{\infty} \frac{d\nu}{1+\nu^2} \frac{\nu^2}{(\nu^2+1/4)^2}
\frac{{\rm tanh} \pi \nu}{\pi \nu} F(\nu) \ {\rm e}^{z \chi(\nu)}
\label{eq:BFKL}
\end{equation}
where
\begin{equation}
z \equiv \frac{3 \alpha_s}{\pi} \log \frac{\hat{s}}{p_t^2},
\end{equation}
$\chi(\nu) = 2(\Psi(1) - {\rm Re}\Psi(1/2+i \nu))$ is the BFKL eigenfunction
\cite{BFKL},
$F(\nu) = 2(11+12 \nu^2)$ for on-shell photons and there is a sum over the
quark charges squared, $e_q^2$. The separation in rapidity between the
struck parton and the final-state photon is 
$\deta \approx \log(\hat{s}/p_t^2).$

The full photon-proton 
cross-section is obtained after multiplying by the parton density functions: 
\begin{equation}
\frac{d \sigma}{dx dp_t^2} = \left[ \frac{81}{16} g(x,\mu) + \Sigma(x,\mu) 
\right] \frac{d \sigma_{\gamma q}}{dp_t^2} \label{exact}
\end{equation}
and we take the factorization scale $\mu = p_t$.

We have implemented this result in the HERWIG event generator in order to
aid the experimental measurement of the process. We also note that having
done this, it is a straightforward procedure to include the high-$p_t$
production of vector mesons in a similar manner and we intend to do this
in the near future. In this paper, we compare the HERWIG generated data with
theory and discuss the strategy for a future measurement.

In order to speed up the event generation procedure, we used the following
approximate parameterization:
\begin{eqnarray}
G(z) &\equiv& \int_{-\infty}^{\infty} \frac{d\nu}{1+\nu^2}
\frac{\nu^2}{(\nu^2+1/4)^2} \frac{{\rm tanh}\pi \nu}{\pi \nu}
2 (11+12 \nu^2) \ \e^{z \chi(\nu)}  \nonumber \\
&\approx& \frac{4.52}{(z+0.1)^{3/2}} \e^{4z \ln 2} \ \Theta(z-1) +
(23.7 + 35 z^{2.3}) \Theta(1-z)
\end{eqnarray}  
which is good to within a few percent over the $z$-range of interest.

\section{Results}

\begin{figure}[ht]
\begin{minipage}[t]{0.475\textwidth}
\centerline{\resizebox{7.5cm}{!}{\rotatebox{0}{\includegraphics{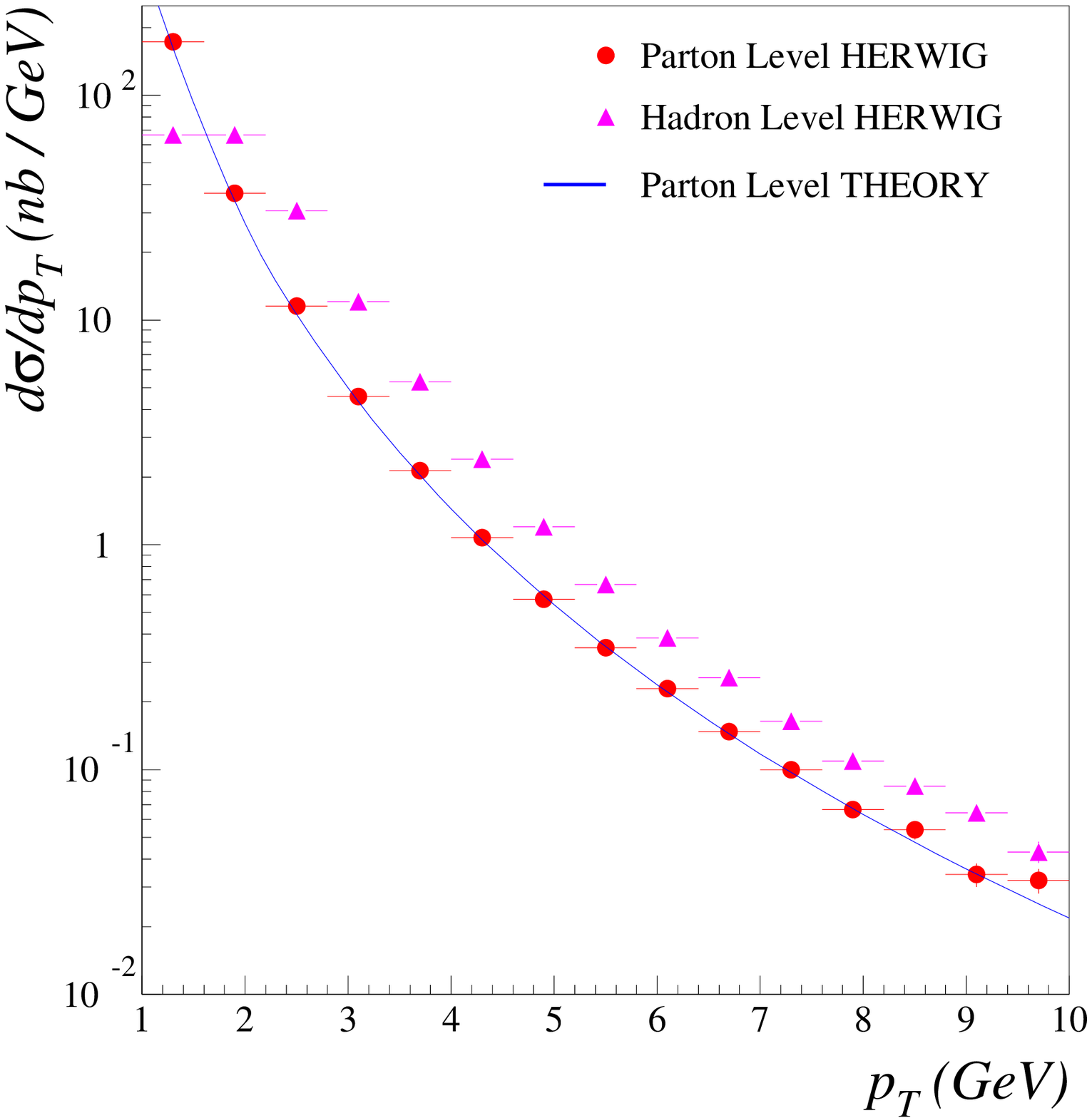}}}}
\caption{\\The photon $p_t$ spectrum.}
\label{ptspec}
\end{minipage}\hspace*{\fill}
\begin{minipage}[t]{0.475\textwidth}
\centerline{\resizebox{7.5cm}{!}{\rotatebox{0}{\includegraphics{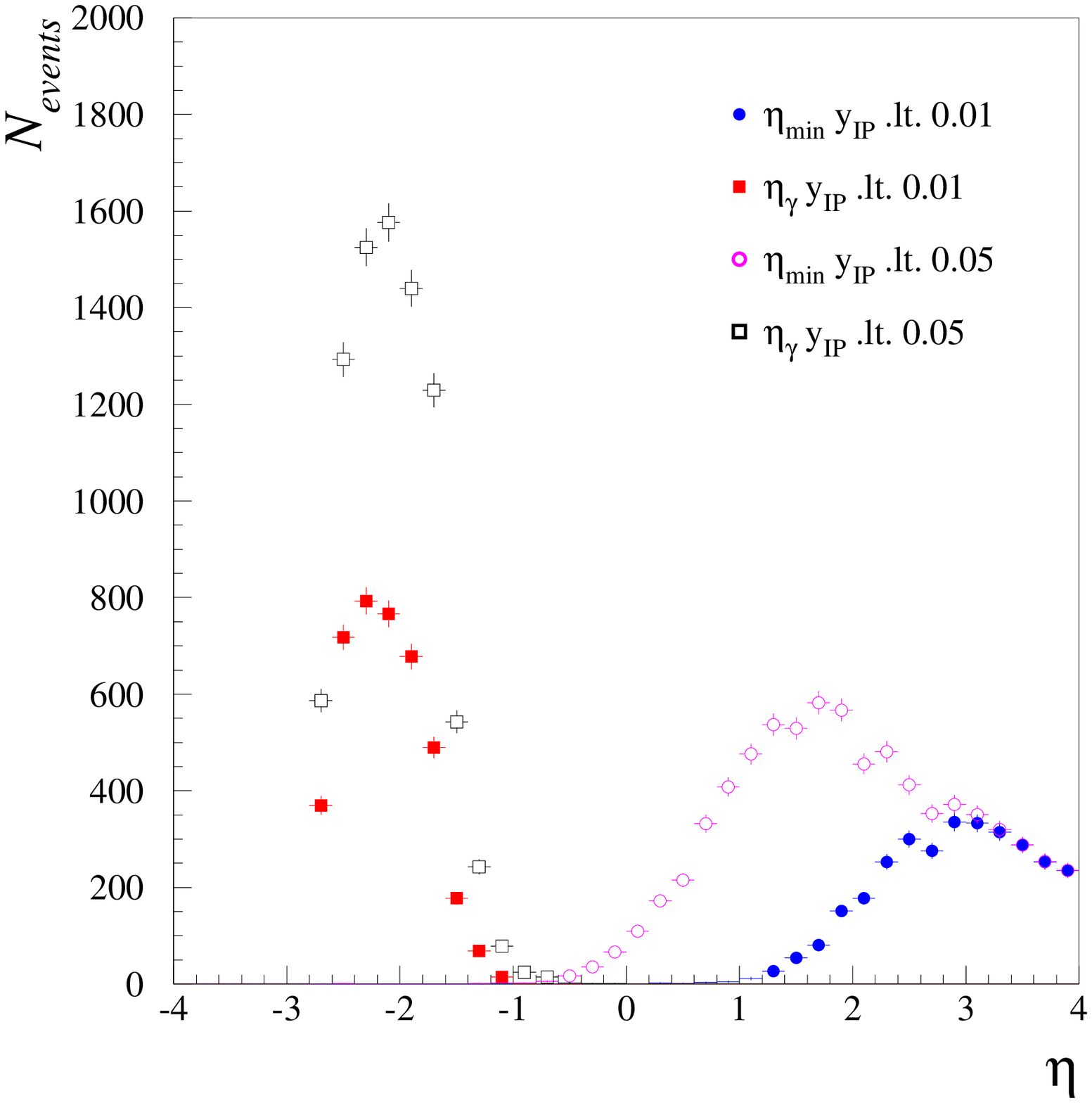}}}}
\caption{\\The rapidities of the scattered photon and edge of system $Y$.}
\label{rap}
\end{minipage}
\end{figure}

We show the $p_t$ spectrum of the scattered photon in Figure \ref{ptspec}.
This plot is computed at fixed $W=200$ GeV ($W^2 = \hat{s}/x$) and 
a fixed $\alps = 0.2$. The choice to fix $\alps$ is supported by Tevatron 
and HERA data on gaps between jets and high $p_t$ diffractive vector 
meson production \cite{alps}. 
The solid curve shows the theoretical prediction derived directly from
(\ref{exact}), it is compared to the HERWIG generated data at the parton
level (before parton showers) and at the hadron level. We used the proton
parton density functions of \cite{GRV} (code 155 in PDFLIB \cite{PDF}).
The good agreement
between theory and parton level is a check that the process is correctly
implemented in HERWIG. The systematic shift arises because HERWIG ensures
that energy and momentum are conserved and that final state hadrons are on
shell \cite{Cox}.

In subsequent figures, we make the typical HERA cuts on the photon energy 
variable, $0.25 < y < 0.75$, and on the photon virtuality, 
$Q^2 < 0.01$ GeV$^2$.
Statistical errors are shown corresponding to 44 pb$^{-1}$ of ep data, typical
of that already collected by each HERA experiment.
We also make a cut $y_{\pom} < 0.01$ where 
\begin{equation}
y_{\pom} = \sum_{i} \frac{(E-p_z)_i}{2 E_{\gamma}} \approx \frac{p_t^2}{x W^2}
\approx \e^{-\deta} 
\end{equation}
and the sum is over all final state particles excluding the electron and 
photon. Note that $y_{\pom}$ can be measured accurately without needing to 
see the whole
of system $Y$ since the material lost at low angles does not contribute 
much to the numerator. As the last approximate equality shows, this cut 
ensures that the rapidity gap between the outoing struck parton and the
outoing photon is bigger than about 4.5 units (recall that a large rapidity
gap is a signal of diffractive processes). We have also integrated over 
all photon $p_t > 2.5$~GeV (and used $p_{{\rm tmin}} = 1$~GeV in the event 
generation).

The effect of varying the $y_{\pom}$ cut on the size 
of rapidity gap can be seen 
in Figure~\ref{rap}, where we plot the rapidity of the scattered 
photon and the edge of system $Y$. By not requiring to measure system $Y$ in 
detail it is possible to reach very high rapidity gaps. As the plot shows, 
gaps of 6 to 7 units in rapidity are not uncommon. Note that the limited
$y$-range combined with the steeply falling $p_t$ spectrum constrain the
rapidity of the photon to be around $\eta \approx -2$. 

\begin{figure}[ht]
\begin{minipage}[t]{0.475\textwidth}
\centerline{\resizebox{7.5cm}{!}{\rotatebox{0}{\includegraphics{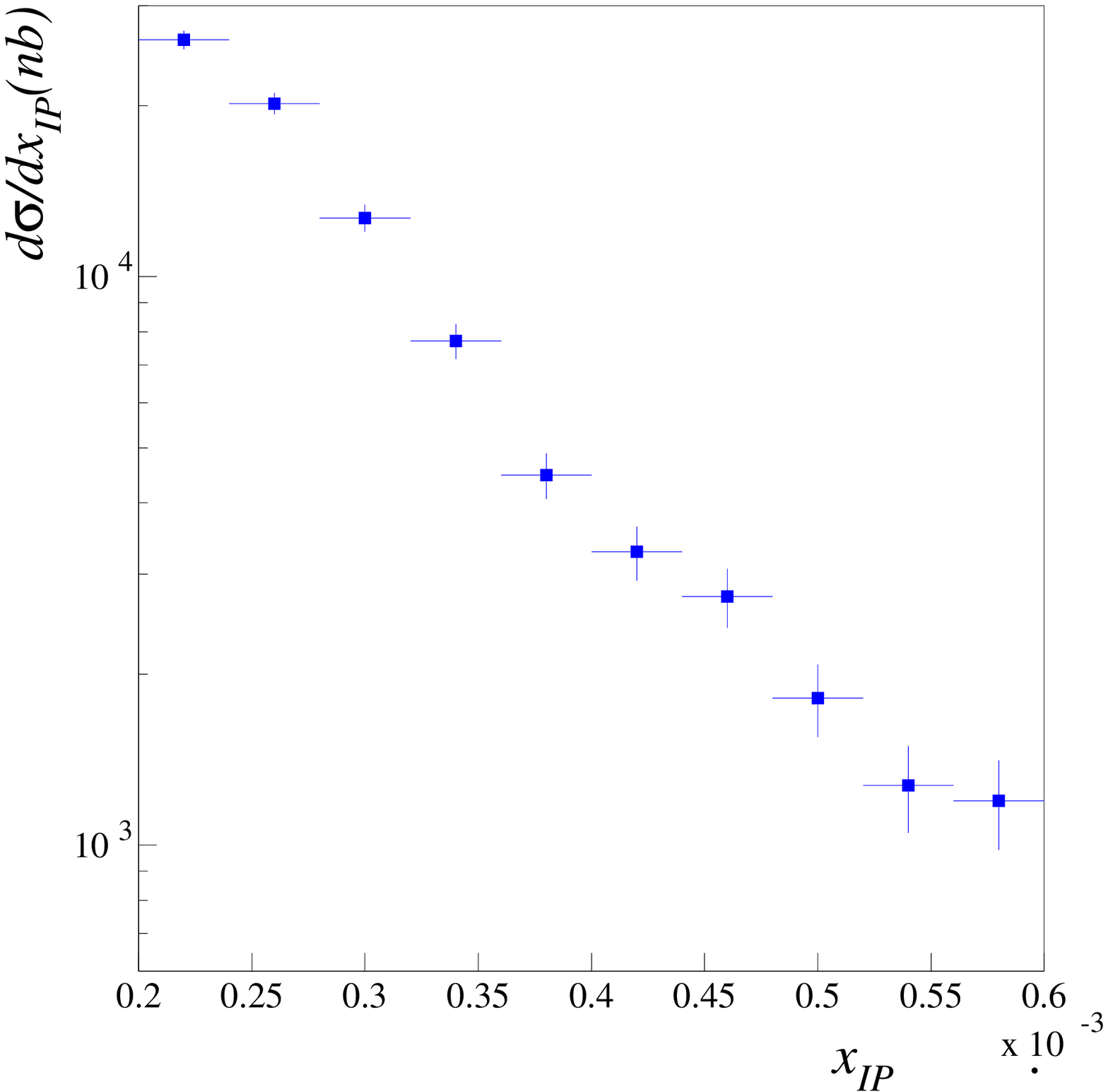}}}}
\caption{\\The $x_{\pom}$ distribution.}
\label{xpom}
\end{minipage}\hspace*{\fill}
\begin{minipage}[t]{0.475\textwidth}
\centerline{\resizebox{7.5cm}{!}{\rotatebox{0}{\includegraphics{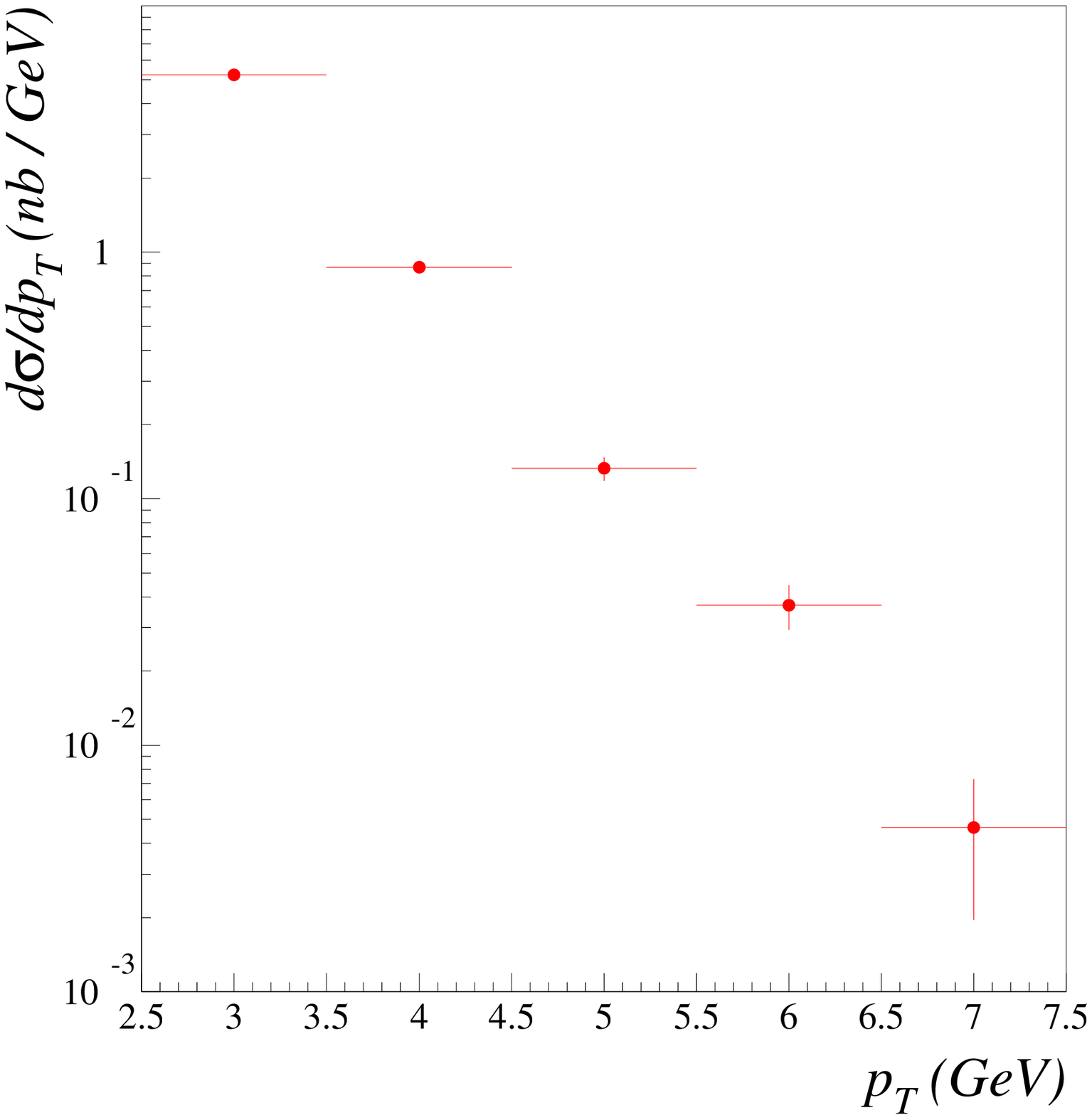}}}}
\caption{\\The $p_t$ distribution.}
\label{pt2}
\end{minipage}
\end{figure}

In Figure \ref{xpom} we show the $x_{\pom}$ distribution and in Figure
\ref{pt2} the $p_t$ distribution:
\begin{equation}
x_{\pom} = \frac{(E+p_z)_{\gamma}}{2 E_p} \approx \frac{p_t^2}{W^2}. 
\end{equation}
This variable can be measured to high accuracy. The steep rise at small
$x_{\pom}$ is driven by the BFKL kernel $\chi(\nu)$ in (\ref{eq:BFKL}). In 
particular, the dominant conribution comes from $\nu \approx 0$, and this
leads to
\begin{equation}
\frac{d \sigma}{d x_{\pom}} \sim \frac{W^2}{p_t^4} \e^{2z \chi(0)}
\sim \frac{1}{W^2} \left( \frac{1}{x_{\pom}} \right)^{2 \omega_0 + 2}
\end{equation}
where $\omega_0 = (3 \alps/\pi) 4 \ln 2$ in the LLA. It will be interesting
to see to what degree the measured $x_{\pom}$ distribution follows this 
power-like behaviour.

\ack
We thank Jon Butterworth and Norman Evanson for their help.


\section*{References}
\begin{thebibliography}{99}
\bibitem{Ivanov1} Ginzburg I F and Ivanov D Yu 1996 \PR {\bf D54} 5523
\bibitem{Ivanov2} Ivanov D Yu and W\"usthoff M 1999 \EJP {\bf C8} 107
\bibitem{Evanson} Evanson N G and Forshaw J R 1999 \PR {\bf D60} 034016
\bibitem{Ginzburg} Ginzburg I F, Panfil S L and Serbo V G 1987 \NP 
{\bf B284} 685
\bibitem{BFKL} Balistky I and Lipatov L N 1978 {\it Sov. J. Nucl. Phys.} 
{\bf 28} 822
\bibitem{alps} Cox B E, Forshaw J R and L\"onnblad L 1999 {\it JHEP} {\bf 10} 
023
\\ Forshaw J R, 1999 \NP (Proc. Suppl.) {\bf 79} 311
\bibitem{GRV} Gl\"uck M, Reya E and Vogt A 1995 \ZP {\bf C67} 433 
\bibitem{PDF} Plothow-Besch H, W5051 PDFLIB, 1997.07.02, CERN-PPE
\bibitem{Cox} Cox B E and Forshaw J R 1998 \PL {\bf B434} 133

\end{thebibliography}
\end{document}